\documentstyle[aps,epsf,tighten,epsfig,amsmath,multicol]{revtex}
\renewcommand{\narrowtext}{\noindent\begin{multicols}{2}\noindent
\global\columnwidth20.5pc}
\renewcommand{\widetext}{\end{multicols}
\global\columnwidth42.5pc}  
\def\bi{{\bf i}}
\def\bj{{\bf j}}

\def\b0{{\bf 0}}
\def\cG{{\cal G}}

\def\cT{{\cal T}}

\def\bra{\langle}
\def\ket{\rangle}
\def\up{\uparrow}
\def\down{\downarrow}

\def\eps{\epsilon}

\def\om{\omega}
\def\sg{\sigma}
\def\Sg{\Sigma}

\begin{document}
\title{Dynamical mean-field theory for pairing and spin gap \\ 
       in the attractive Hubbard model}
\author{M.\ Keller$^1$, W.\ Metzner$^1$, and U.\ Schollw\"ock$^2$ \\
{\em $^1$Theoretische Physik C, Technische Hochschule Aachen, 
 D-52056 Aachen, Germany} \\
{\em $^2$Sektion Physik, Universit\"at M\"unchen, D-80333 M\"unchen,
 Germany}}
\date{\small\today}
\maketitle
\begin{abstract}
We solve the attractive Hubbard model for arbitrary interaction
strengths within dynamical mean-field theory.
We compute the transition temperature for superconductivity
and analyze electron pairing in the normal phase.
The normal state is a Fermi liquid at weak coupling and a non-Fermi 
liquid state with a spin gap at strong coupling.
Away from half-filling, the quasi-particle weight vanishes 
discontinuously at the transition between the two normal states.
\noindent
\mbox{PACS: 71.10.Fd, 71.10.-w, 74.20.Mn} \\
\end{abstract}

\narrowtext


Fermi systems with a weak attractive interaction are Fermi liquids
which undergo a phase transition into a superconducting state via
the condensation of weakly bound Cooper pairs at a low critical
temperature $T_c$. 
For a long time this weak coupling route to superconductivity, which 
is well described by the BCS mean-field theory \cite{BCS}, was 
distinguished almost dogmatically from Bose-Einstein condensation
of bosons, although in both cases a phase transition characterized 
by the same $U(1)$ symmetry breaking occurs.
Leggett \cite{Leg} was the first to point out that BCS 
superconductivity can be continuously connected to Bose-Einstein
condensation of tightly bound pairs by increasing the two-particle
attraction between the fermions from weak to strong coupling.
The size of the Cooper pairs shrinks continuously until spatially
well separated bosons form, which undergo Bose condensation at a
sufficiently low temperature.
Nozi\`eres and Schmitt-Rink \cite{NS} have extended Leggett's
analysis to lattice electrons and finite temperatures. 
Based on physical insight gained from a discussion of the weak 
and strong coupling limits and an approximate (T-matrix) treatment 
of the intermediate regime they concluded that the evolution from 
weak coupling to strong coupling superconductivity is indeed 
smooth.
\par
The interest in the intermediate regime between the BCS and 
Bose-Einstein limits increased considerably after the discovery of 
high-temperature superconductors, which are characterized by Cooper 
pairs whose size is only slightly bigger than the average electron 
distance \cite{Gin}.
Much recent work has therefore been dedicated to the theory of Fermi 
systems with attractive interactions of arbitrary strength 
\cite{Ran1}.
It was shown that sufficiently strong attraction or reduced 
dimensionality can lead to energy gaps even in the {\em normal}\/ 
phase, which have been related to pseudogap phenomena in the cuprate 
superconductors \cite{Ran2}.
\par
In this work we analyze the formation of pairs in a Fermi system
with arbitrary attractive interactions by solving the attractive
Hubbard model within dynamical mean-field theory (DMFT) \cite{GKKR}.
We show that the normal state is a Fermi liquid at weak coupling
and a non-Fermi liquid state characterized by bound pairs and a
spin gap at strong coupling, in qualitative agreement with Quantum 
Monte Carlo (QMC) studies of the two- and three-dimensional 
attractive Hubbard model \cite{RTMS,SPSBM,San}.
At very low temperatures the transition between the Fermi liquid 
and the normal paired state is discontinuous.
\par


In standard notation the Hubbard model for lattice fermions
with a nearest neighbor hopping amplitude $-t$ and a local 
interaction $U$ is given by
\begin{equation}\label{Hubbard}
 H = -t \sum_{\bra\bi,\,\bj\ket} \sum_{\sg} 
 c^{\dag}_{\bi\sg} \, c_{\bj\sg} 
 \, + \, U \, \sum_{\bj} n_{\bj\up} \, n_{\bj\down} \; .  
\end{equation}
The attractive ($U<0$) Hubbard model is a superconductor below a 
certain critical temperature $T_c(U,n) > 0$ for all $U$ at any 
average density $n$, if the lattice dimensionality is above two 
\cite{MRR}.
At half-filling ($n=1$) the usual $U(1)$ gauge symmetry becomes 
a subgroup of a larger $SO(3)$ symmetry, and the superconducting
order parameter mixes with charge density order.
In two dimensions one expects a Kosterlitz-Thouless phase at low 
temperatures for all $U < 0$ and $n \neq 1$ \cite{MRR}. 
\par
In the weak coupling limit $U \to 0$ and dimensions $d > 2$ the 
attractive Hubbard model can be treated by BCS mean-field 
theory \cite{NS,MRR}.
In the strong coupling limit $U \to -\infty$ the low energy 
sector of the model (excitation energies $\ll |U|$) can be 
mapped onto an effective model of hard core lattice bosons with
a hopping amplitude of order $t^2/U$ and a repulsive nearest 
neighbor interaction of the same order \cite{NS,MRR}. 
These bosons undergo Bose condensation in $d>2$ dimensions and
a Kosterlitz-Thouless transition in two dimensions (for $n \neq 1$) 
at a critical temperature of order $t^2/|U|$.
\par
For nearest neighbor hopping on a bipartite lattice the particle-hole 
transformation of spin-$\up$ fermions
\begin{equation}\label{phtrafo}
 c_{\bj\up} \mapsto \eta_{\bj} \, c^{\dag}_{\bj\up} \; , \quad 
 c^{\dag}_{\bj\up} \mapsto \eta_{\bj} \, c_{\bj\up} \, ,
\end{equation}
where $\eta_{\bj} = 1 \; (-1)$ for $\bj$ on the A-sublattice
(B-sublattice), maps the attractive Hubbard model at density
$n$ onto a repulsive Hubbard model at half-filling with a
finite average magnetization $m = 1-n$ \cite{MRR}.
We will use this relation to compare with results known for the 
repulsive Hubbard model.
\par


We have solved the attractive Hubbard model within DMFT \cite{GKKR}.
In contrast to other (simpler) mean-field approaches, DMFT provides 
an exact solution of the model in the limit of infinite lattice 
dimensionality \cite{MV}, since it captures local fluctuations
exactly. 
\par
To convince ourselves that DMFT is a suitable approach for the
weak to strong coupling crossover problem, let us first consider
the limits.
At weak coupling DMFT captures the complete BCS physics, since it 
contains the Feynman diagrams contributing to the BCS mean-field 
theory. 
At strong coupling, where the attractive Hubbard model maps to the
hard core Bose gas, DMFT reduces to the standard mean-field theory 
of the hard core Bose gas \cite{fn1}.
Hence, Bose-Einstein condensation of preformed pairs is obtained at 
a critical temperature of order $t^2/|U|$ at large $|U|$.
\par
Within DMFT the fluctuating environment of any lattice site is
replaced by a local but dynamical effective field $\cG_0$ 
\cite{GKKR}.
The mean-field equations involve the calculation of the propagator
$G(\tau) = - \bra \cT c_{\sg}(\tau) c^{\dag}_{\sg}(0) \ket$ of an
effective single-site Hubbard model coupled to the dynamical field 
$\cG_0$, and a self-consistency condition relating $G$ to the local 
propagator of the full lattice system. 
The lattice structure enters only via the bare density of
states (DOS), as long as the translation invariance of the lattice 
is not broken. 
We have used the particularly simple self-consistency equations 
\cite{GKKR}
\begin{equation}
 \cG_0^{-1}(i\omega) = i\omega + \mu - (\eps_0/2)^2 G(i\omega)
\end{equation}
valid for a half-ellipse shaped density of states 
$D_0(\eps) = \frac{2}{\pi\eps_0^2} \sqrt{\eps_0^2-\eps^2}$.
Any other bounded DOS would yield qualitatively similar results.
In the following we will set $\eps_0/2 = 1$.
Susceptibilities such as the pairing and the spin susceptibility 
can also be computed from expectation values of operator products 
within the effective single-site problem. 
The DMFT equations can also easily be extended to superconducting
or other symmetry broken phases \cite{GKKR}.
In this work we focus however on normal state properties.  
\par
The effective single-site problem cannot be solved analytically.
We have solved it numerically by discretizing the imaginary time 
interval and computing expectation values via the standard 
Hirsch-Fye algorithm \cite{HF}. 
\par


We now present and discuss results from our DMFT calculation
at quarter-filling ($n=1/2$). 
We do not expect that the results depend qualitatively on the
density in the attractive Hubbard model, as long as $n$ is finite.
Only the particle-hole symmetric case $n=1$ is special due to its 
larger symmetry group.
Quarter-filling is well below half-filling but still high enough 
to see collective many-body effects, which are not obtained in the 
low-density limit.
\par
The critical temperature $T_c$ for the onset of superconductivity
shown in Fig.\ 1 has been obtained from the pairing susceptibility
in the normal phase, which diverges as the temperature approaches
$T_c$ from above.
The critical temperature at half-filling, where the superconducting
state is degenerate with a charge-density wave state, has been 
computed within DMFT already earlier \cite{FJS}. 
Correlations of course suppress $T_c$ with respect to the BCS
result.
At strong coupling $T_c(U)$ approaches the $1/U$ behavior described
by the hard core Bose gas limit.
At intermediate coupling strengths, $T_c(U)$ varies smoothly, as
predicted already by Nozi\`eres and Schmitt-Rink \cite{NS}.
A comparison with a recent result for $T_c$ obtained by combining
the DMFT with a self-consistent T-matrix approximation (TMA) 
\cite{KMS} shows that the latter approximation reproduces the correct 
qualitative behavior of $T_c(U)$, but fails quantitatively.
\begin{figure}
\vskip 6mm
\begin{center}
\epsfig{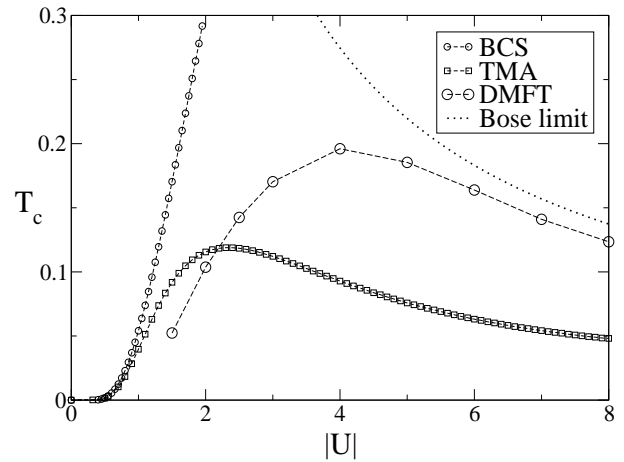} 
\end{center}
\caption{\label{fig1}{Critical temperature $T_c$ as a function of the 
coupling strength $|U|$ at quarter-filling within DMFT, compared to 
$T_c$ obtained from BCS theory and from the T-matrix approximation 
(TMA), respectively.}}
\end{figure}
\par
In the following we discuss the weak to strong coupling crossover
in the {\em normal}\/ phase. 
We ignore the superconducting instability and study normal state
solutions of the DMFT equations also below $T_c$. 
Of course these solutions do not minimize the free energy, but
they could be stabilized by the field energy of a sufficiently
strong external magnetic field.
\par
At weak coupling the normal state of the system is a Fermi liquid
with fermionic quasi-particle excitations. Besides numerical
evidence this follows \cite{fn2} from the analyticity of weak 
coupling perturbation theory for the effective single-site 
problem. 
By contrast, at sufficiently strong coupling $|U| \gg \eps_0$ 
and zero temperature all particles are bound in pairs, because a
small kinetic energy cannot overcome a finite binding energy. 
Only short-ranged virtual breaking of local pairs occurs.
At low finite temperatures $T \ll |U|$ only an exponentially 
small fraction of pairs dissociates.
\par
A good measure for local pair formation is the density of doubly 
occupied sites $n_d = \bra n_{\bj\up} \, n_{\bj\down} \ket \:$.
For an uncorrelated state the density of doubly occupied sites
is simply the product of the average density of up and down spin 
particles, i.e.\ $n_d^0 = n_{\up} \, n_{\down} = (n/2)^2$.
An attractive interaction enhances $n_d$.  
In the limit of infinite attraction all particles are bound as 
local pairs such that $n_d \to n/2$. 
In Fig.\ 2 we show results for $n_d(T)$ for various $U$.
\begin{figure}
\vskip 6.5mm
\begin{center}
\epsfig{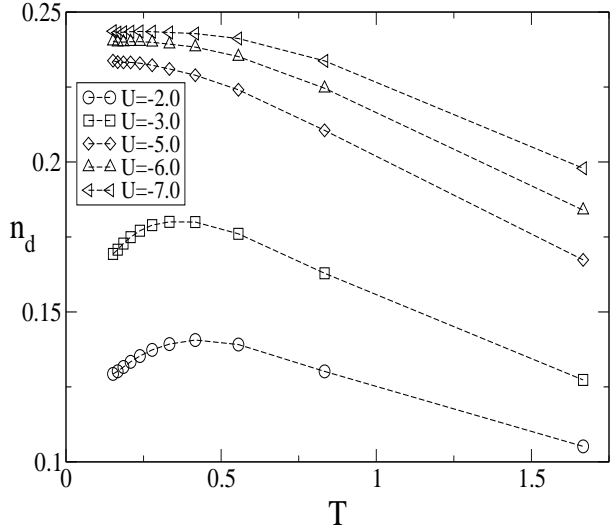} 
\end{center}
\caption{\label{fig2}{Density of local pairs $n_d$ as a function of 
 temperature for various coupling strengths $U$ at quarter-filling.}}
\end{figure}
\noindent
For $T \to \infty$ the density of doubly occupied sites tends to
$(n/2)^2$, corresponding to an uncorrelated state.
For decreasing temperature $n_d(T)$ first increases as a 
consequence of the attractive interaction. 
For small or moderate $U$, however, $n_d(T)$ slightly decreases 
again at low temperatures. This effect, which has also been 
obtained in the combined DMFT + TMA calculation \cite{KMS}, can 
be attributed to the kinetic energy, which tends to dissociate 
pairs if the attraction is not too strong.
Note that in the pairing regime for stronger $U$ the upturn in
$n_d(T)$ at low temperatures is missing. The kinetic energy is
not able to unbind pairs any more.
A completely analogous (particle-hole transformed) behavior has 
been found in the DMFT solution of the repulsive Hubbard model at
half-filling \cite{GKKR}.
\par
The absence of fermionic quasi-particles in the pairing state at
strong coupling also leads to a pronounced {\em spin gap}\/.
In Fig.\ 3 we show our DMFT results for the temperature dependence 
of the spin susceptibility $\chi_s$, for various coupling strengths.
For a weak attraction the spin susceptibility increases
monotonously for lower temperatures and then saturates at a 
finite value for $T \to 0$, as expected for a Fermi liquid.
For strong coupling, however, $\chi_s$ decreases rapidly at low 
temperatures, as expected for a system where spin excitations are 
gapped. 
This gap, which has also been seen in QMC simulations of the
two-dimensional \cite{RTMS,SPSBM} and three-dimensional \cite{San}
Hubbard model, is clearly due to the binding energy of pairs in 
the non-Fermi liquid state forming at strong coupling.
For a moderate attraction, pseudogap behavior seems to set in at 
intermediate temperatures, but for $T \to 0$ the susceptibility 
tends to a reduced but non-zero value.
We expect that this behavior reflects the presence of a narrow
quasi-particle band in the system, equivalent to the one known
for the repulsive model near the Mott transition \cite{GKKR}.
The selfconsistent TMA fails to yield spin gap behavior in high 
dimensions \cite{KMS}, and yields only a rather weak suppression 
of $\chi_s$ in two dimensions \cite{MPSSRB}.
\begin{figure}
\vskip 6mm
\begin{center}
\epsfig{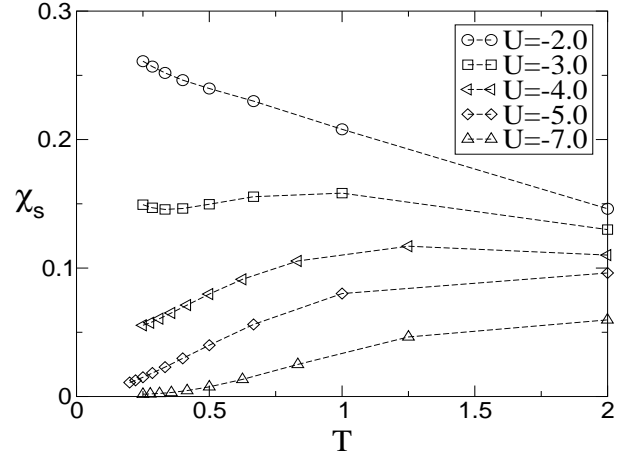} 
\end{center}
\caption{\label{fig3}{Spin susceptibility $\chi_s$ as a function of 
 temperature for various coupling strengths $U$ at quarter-filling.}}
\end{figure}
\par
Since the Fermi liquid state at weak coupling is qualitatively 
different from the bound pair state at strong coupling, there
must be a sharply defined {\em pairing transition}\/ at some
critical attraction $U_c$ at least in the ground state.
At quarter-filling we can estimate from our data 
$U_c \approx -2.5 \eps_0$.
\par
To see how the Fermi liquid breaks down upon increasing the
attraction strength, we have computed the renormalization 
factor $Z(T) = [1 - \Sg(i\omega_0)/i\omega_0]^{-1}$, where $\Sg$ 
is the self-energy and $\om_0 = \pi T$ the smallest (positive) 
Matsubara frequency at temperature $T$.
In Fig.\ 4 we plot $Z(T)$ as a function of $T$ for various 
$U$ at quarter-filling (left) and at half-filling (right). 
At quarter-filling $Z(T)$ extrapolates to a finite positive value $Z$ 
in the limit $T \to 0$, for any $U$.
In the Fermi liquid phase $Z$ has physical meaning, being the 
spectral weight for quasi-particles, the Fermi edge discontinuity 
in the momentum distribution function and, within DMFT, also the 
inverse mass renormalization.
This meaning is of course lost in the bound pair state.
Note that the finiteness of $Z$ for all $U$ does not imply 
that the system is a Fermi liquid for arbitrary interactions!
A simple calculation shows that $Z$ is finite even in the
atomic limit $t=0$, where the system is obviously not a Fermi
liquid.
An exception is the half-filled case, where $Z \to 0$ for 
$U \to U_c$, and $Z=0$ for all $|U| \geq |U_c|$ (see Fig.\ 4) 
and in the atomic limit.
For half-filling the continuous vanishing of $Z$ has been
analyzed in much detail for the Mott transition in the
repulsive Hubbard model \cite{GKKR}, which is equivalent to
the pairing transition in the attractive model by virtue of
the particle-hole symmetry (2).
We thus conclude that the Fermi liquid phase disappears with
a finite $Z$ at electron densities $n \neq 1$, i.e.\ the 
quasi-particle weight disappears {\em discontinuously} at the 
pairing transition.
\vskip 5mm
\begin{figure}
\begin{center}
\epsfig{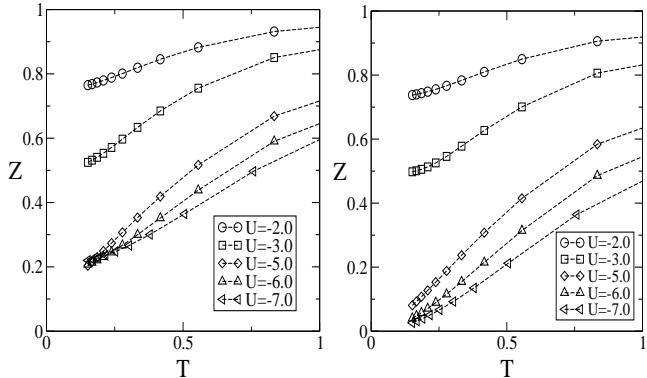} 
\end{center}
\caption{\label{fig4}{$Z(T)$ as a function of temperature for 
 various $U$ at quarter-filling (left) and half-filling (right).}}
\end{figure}
\par
It is instructive to consider the low density limit $n \to 0$ 
for comparison. In that limit the bound pair state is stable
once the attraction exceeds the threshold for two-particle 
binding $U_c^0$. For $|U| < |U_c^0|$ no bound states exist,
and the particles move essentially freely, due to the low
density, and $Z$ is almost one even close to the transition.
\par
Using the Hirsch-Fye algorithm it is hard to determine 
unambiguously whether there is a sharp phase transition at some 
critical $U_c(T)$ also at sufficiently low finite temperature 
or only a smooth (though steep) crossover, since the computation 
time increases rapidly at low temperatures. 
However, one can find an answer to this question by exploiting the 
equivalence of the attractive Hubbard model at generic densities and 
the half-filled repulsive Hubbard model with a finite magnetization. 
The latter model has been analysed earlier within DMFT by Laloux 
et al.\ \cite{LGK}, who solved the DMFT equations with an
exact diagonalization algorithm which is more efficient 
than the Hirsch-Fye algorithm at low temperatures.
The results of their work imply that at very low temperatures a 
first order transition occurs in the attractive Hubbard
model between a thermally excited Fermi liquid state and a 
thermally excited bound pair state.
\par
In summary, we have shown that DMFT yields a transition from a
Fermi liquid state at weak coupling to a non-Fermi liquid state
with a spin-gap at strong coupling in the attractive Hubbard model.
Spin-gap behavior for {\em strong}\/ attraction is obviously governed 
by {\em short-range}\/ pair correlations which are captured by DMFT.
Long-range superconducting fluctuations characteristic for 
low-dimensional systems are crucial for normal state gaps only at 
weaker coupling.

\smallskip
\noindent{\bf Acknowledgments:}
W.M. would like to thank A.\ Georges for his kind hospitality
and valuable discussions at Ecole Normale Superieure.
Discussions with M.\ Randeria and M. Radke are also gratefully 
acknowledged. 
This work has been supported by the Deutsche Forschungsgemeinschaft 
under Contract No.\ Me 1255/5.

\widetext


\begin{thebibliography}{99}
\bibitem{fn1} The hard core Bose gas can be mapped to an 
 antiferromagnetic Heisenberg model in an external magnetic field,
 and its mean-field theory maps to the standard mean-field theory
 of the Heisenberg model; 
 see, for example Ref. \cite{MRR}, Sec.\ IIB.
\end{thebibliography}
\end{document}